\documentclass[%
 reprint,
superscriptaddress,
preprintnumbers,  
nofootinbib,
 amsmath,amssymb,
 aps,
]{revtex4-1}

\usepackage{graphicx}
\usepackage{dcolumn}
\usepackage{bm}
\usepackage{natbib}
\usepackage{siunitx}
\usepackage{amsmath}
\usepackage{amssymb}
\usepackage{amsfonts}
\usepackage{xfrac}
\usepackage{graphicx}
\usepackage{subfigure}
\usepackage{hyperref}
\usepackage[dvipsnames]{xcolor}
\usepackage[colorinlistoftodos]{todonotes}
\usepackage[normalem]{ulem}

\definecolor{navyblue}{rgb}{0.0, 0.0, 0.5}
\definecolor{royalblue}{rgb}{0.25, 0.41, 0.88}
\definecolor{cadmiumgreen}{rgb}{0.0, 0.42, 0.24}
\definecolor{blue-violet}{rgb}{0.54, 0.17, 0.89}
\definecolor{darkviolet}{rgb}{0.58, 0.0, 0.83}
\definecolor{orange(colorwheel)}{rgb}{1.0, 0.5, 0.0}

\usepackage{hyperref}
\hypersetup{
    colorlinks=true, 
    linkcolor=royalblue, 
    citecolor=magenta}

\newcommand\be{\begin{equation}}
\newcommand\ee{\end{equation}}
\newcommand\bea{\begin{eqnarray}}
\newcommand\eea{\end{eqnarray}}

\newcommand{\lcdm}{\Lambda\mathrm{CDM}}

\newcommand\ie{{\it i.e.}~}

\usepackage{booktabs}
\usepackage{multirow}
\usepackage{dcolumn}
\usepackage{colortbl}
\newcommand\vertsp{\rule[-2mm]{1mm}{0mm} &}
\newcommand\horsp{\rule[-1.5mm]{0mm}{4.125mm}}
\newcommand\morehorsp{\rule[-2.25mm]{0mm}{6mm}}

\definecolor{magenta(process)}{rgb}{1.0, 0.0, 0.56}

\definecolor{darkspringgreen}{rgb}{0.09, 0.45, 0.27}

\definecolor{royalblue(web)}{rgb}{0.25, 0.41, 0.88}

\begin{document}

\title{Current and future constraints on single-field $\alpha$-Attractor models }

\author{Guadalupe Cañas-Herrera}
\email{canasherrera@lorentz.leidenuniv.nl}
\affiliation{%
Leiden Observatory, Leiden University, PO Box 9506, Leiden 2300 RA, The Netherlands}
\affiliation{%
Lorentz Institute for Theoretical Physics, Leiden University, PO Box 9506, Leiden 2300 RA, The Netherlands
}%

\author{Fabrizio Renzi}
\email{renzi@lorentz.leidenuniv.nl}
\affiliation{%
Lorentz Institute for Theoretical Physics, Leiden University, PO Box 9506, Leiden 2300 RA, The Netherlands
}%

\date{\today}

\begin{abstract}
We study here the observational constraints on single-field inflationary models achievable with the next generation of CMB experiments. We particularly focus on a Stage-IV like experiment and forecasts its constraints on inflationary parameters in the context of $\alpha$-attractor inflation comprising a large class of single-field models. To tailor our forecasts we use as a fiducial model the results obtained with current CMB and LSS data, assuming the $\alpha$-model a priori. We found that current CMB data are able to place a tight bound on the ratio of the tensor-to-scalar ratio with the alpha parameter $r/\alpha = 3.87^{+0.78}_{-0.94}\cdot 10^{-3}$ and on the running of the scalar index $\alpha_S = -6.4^{+1.6}_{-1.3}\cdot 10^{-4}$ with a value of the scalar index consistent with current constraints. 
In the optimistic scenario of detection of primordial gravitational waves in the CMB B-mode polarization power spectra, we found that CMB-S4 will be able to achieve a $15\%$ bound on the value of the parameter $\alpha = 1.01^{+0.14}_{-0.18}$. This bound clearly show the ability of CMB-S4 to constrain not only the energy scale of inflation but also the shape of its potential. Enlarging the baseline model to also include the neutrino sector merely reduce the accuracy of $5\%$ leading to $\alpha = 1.07^{+0.18}_{-0.23}$ so that our main conclusions are still valid. 
\end{abstract}

\maketitle

\section{Introduction}
In this letter, we forecast the possible constraints that a future CMB StageIV (CMB-S4 hereafter) experiment may impose on inflationary observables in the optimistic scenario of a detection of non-vanishing tensor anisotropies in the Cosmic Microwave Background (CMB) polarization and temperature data. In general, the approach followed within the community (see e.g. \cite{Planck2013_inflat,Planck2015_inflat,BICEP,BICEP2,Aghanim:2018eyx}) is to sample the inflationary parameters without assuming any specific inflationary model a priori. While this approach has the advantage of exploring the inflationary sector model-independently, it does not allow for a complete sampling study of the parameter space in a specific model. Moreover, the assumption that the inflationary observables are independent of one another is in contrast with the prediction of any theory of inflation, which, for instance, assumes the validity of the slow-roll conditions (see e.g. \cite{Renzi:2019ewp,Shokri:2019rfi, PhysRevD.99.123522}). In this work, conversely to the current literature on the subject, we follow a model-dependent approach imposing a specific model \emph{a priori} and calculate the inflationary observables directly imposing the slow-roll conditions on the inflationary potential. 

The Standard $\Lambda$CDM model is based on the simplest inflationary paradigm: canonical slow-roll single-field inflation. Inflation does not only solve the need of fine-tuning the initial conditions from the hot Big Bag scenario, but it also provides an elegant mechanism to explain the origin of the scalar primordial perturbations that evolved into the current cosmic structures at large scales. Furthermore, quantum inflationary fluctuations are expected to source a stochastic background of gravitational waves - so-called Primordial Gravitational waves (PGWs) - sourcing fluctuations in the polarization of CMB photons at recombination leading to a very distinctive signature in the CMB B-modes power spectrum at large angular scales. 
Within this approach, the power spectra of scalar and tensor comoving curvature perturbations are parametrised as power laws:
\begin{align}
    P_S(k) &= A_S \bigg(\frac{k}{k_\star^S}\bigg)^{n_S-1 + \frac{\alpha_S}{2}\log{k/k^S_\star}}  \\
    P_T(k) &= r A_S \bigg(\frac{k}{k_\star^T}\bigg)^{n_T}
\end{align}
where $k^S_\star = 0.05\ \text{Mpc}^{-1}$ and $k^T_\star = 0.002\ \text{Mpc}^{-1}$ and the subscripts stand for scalar and tensor perturbations respectively. The powers of the parametrizations are the scalar and tensor indices ($n_S$ and $n_T$) and the running of the spectral index $\alpha_S$. Statistical analysis of recent cosmological observations (Planck observations of the Cosmic Microwave Background (CMB) \cite{Akrami:2018odb,Aghanim:2018oex,Aghanim:2018eyx} and Large Scale Structure (LSS) surveys \cite{To:2020bhf, 2020arXiv200715632H}) support this parametrisation for the scalar fluctuations with $10^{9}A_S \approx 2.1 $ and $n_S \approx 0.965$ \cite{Akrami:2018odb,Aghanim:2018oex,Aghanim:2018eyx}.

In the last decade, the bound on the amplitude of PGWs (parametrized typically with the tensor-to-scalar ratio, $r$) has not yet seen significant improvement, where only an upper limit $r_{0.002} < 0.056$ at $95\%$ C.L. has been provided in the last data release of the Planck Collaboration \cite{Akrami:2018odb} combining Planck and BICEP2/Keck array (BK15) data \cite{Ade:2018gkx}. Thus, direct detection of tensor modes is still missing. 
Detecting those PGWs would give a direct measurement of the energy scale during inflation, as well as a clear distinguishable signature of the quantum origin of primordial fluctuations. In the upcoming decade, a new generation of CMB experiments (e.g. BICEP3 \cite{BICEP3}, CLASS \cite{CLASS} , SPT-3G \cite{SPT-3G}, Advanced ACTPol \cite{ACTPol}, LBIRD \cite{LBIRD} and CMB-S4 \cite{CMB-S4}) are expected to strongly improve the sensitivity on the B-modes polarization in the Cosmic Microwave Background (CMB), possibly revealing first evidences for inflationary tensor modes with amplitudes $r \sim 0.01 - 0.001$. That range is precisely expected in many well-motivated models, such as the Starobinsky inflation, which is considered the benchmark of future CMB experiments. 
However, while a measure of a non-vanishing $r$ would be of key importance for inflationary theories, it will not allow understanding the inflationary mechanism in detail but only its energy scale. It is therefore timely to investigate, given future CMB experiments, what would be the freedom in a generic inflationary framework that is left in case of the optimistic scenario of a non-vanishing tensor-to-scalar ratio measure. 

In particular, there is a general class of models called  $\alpha$-attractors, that has gained lots of popularity because of their agreement with observational constraints and the universality of their predictions for the inflationary observables \cite{Iarygina:2018kee,Iarygina:2020dwe}. This set of models have also been embedded in a more general multi-field inflationary scenario and in $\mathcal{N} = 1 $ supergravity. In the context of supergravity, the $\alpha$-attractor can be represented by a potential of the form:

\begin{equation}\label{eq.alpha_pot}
    \frac{V(\varphi)}{V_0}= \left(\tanh(\beta\varphi/2)\right)^{2n}
\end{equation}
where $\beta^2 = 2/3\alpha $ and $n$ is an arbitrary value. It is important to note that the "attractor behaviour" of this potential sits on the fact that the observable predictions are the same up to leading order regardless of the value of $n$ while they differs only in sub-leading corrections. 
Assuming slow-roll inflation and the $\alpha$-attractor form of the inflationary potential the observational predictions for the inflationary observables can be written as:  
\begin{subequations}\label{eq.infparameters_obs}
\begin{equation}\label{eq:N}
    r = \frac{12\alpha}{N^2}
\end{equation}
\begin{equation}\label{eq:n_s}
    n_S = 1-\frac{2}{N} = 1 - \sqrt{\frac{r}{3\alpha}},
\end{equation}
\begin{equation}\label{eq:n_run}
    \alpha_S = -\frac{2}{N^2} = -\frac{r}{6\alpha}.
\end{equation}
\end{subequations}
where $N$ is the number of e-folds to inflation to last. These definitions in terms of parameter $\alpha$ encompass several inflationary model and clearly reduces to the well-know Starobinsky inflation for $\alpha = 1$ \cite{Kallosh:2013yoa,Kallosh:2017wnt,Carrasco:2015uma}. Moreover, for a broad class of potentials $V$, as long as $\alpha \ll O(1)$, the scalar spectral index $n_S$, its running $\alpha_S$ and the tensor-to-scalar ratio $r$ converge to the functional form of \autoref{eq.infparameters_obs} regardless of the kinectic terms of the theory \cite{Achucarro:2017ing}. Furtheremore, it has been showed that this statement holds true in some multi-field inflation regimes \cite{Achucarro:2017ing}, where the conditions that guarantee the universality of the observational predictions for the inflationary parameters are derived by imposing constraints on the potential.
This is one of the most important features of single-field $\alpha$-attractor models: the observational predictions for the inflationary observables are universal.
\begin{figure*}[htp!]
    \centering
    \includegraphics[width=0.49\linewidth]{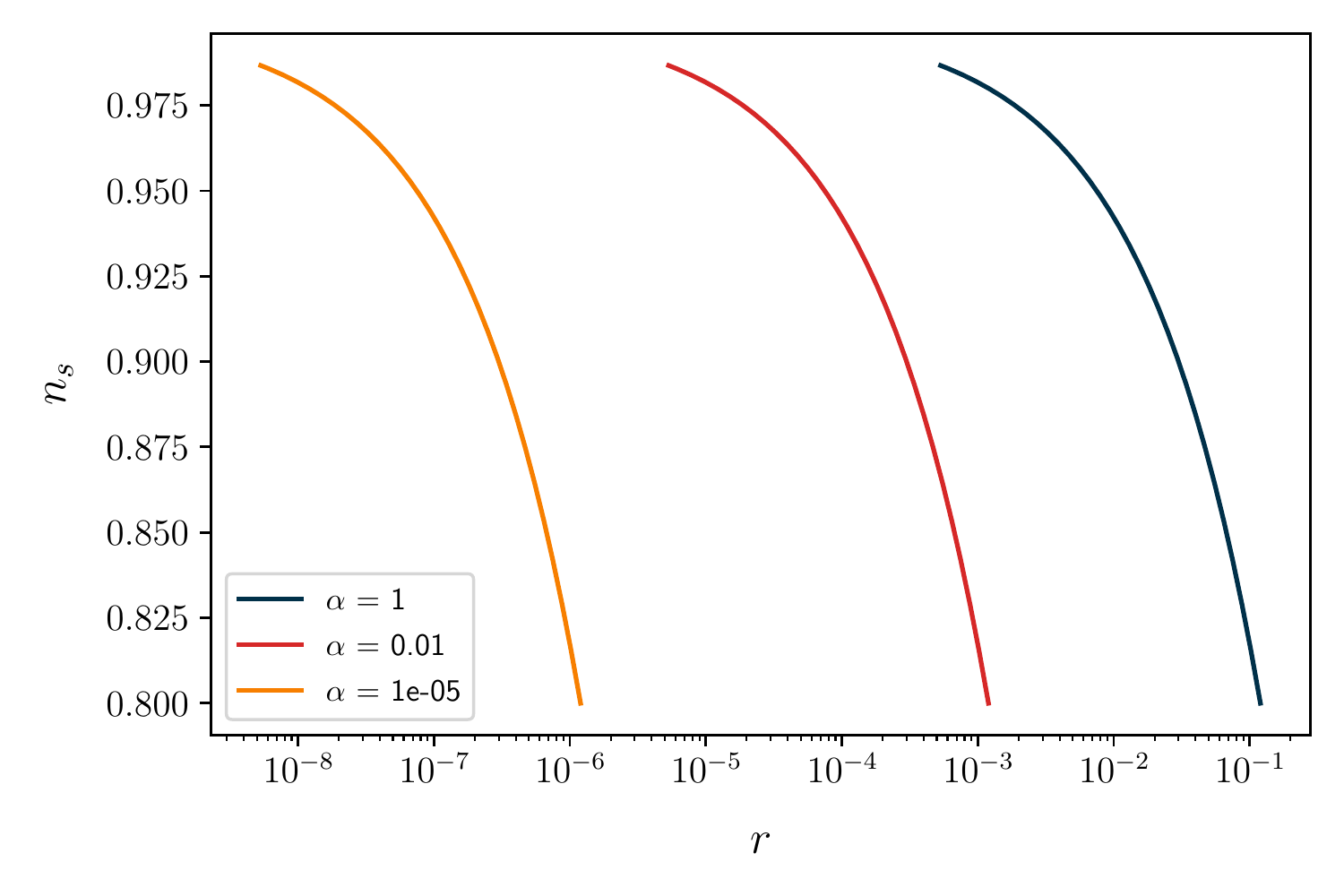}
    \includegraphics[width=0.49\linewidth]{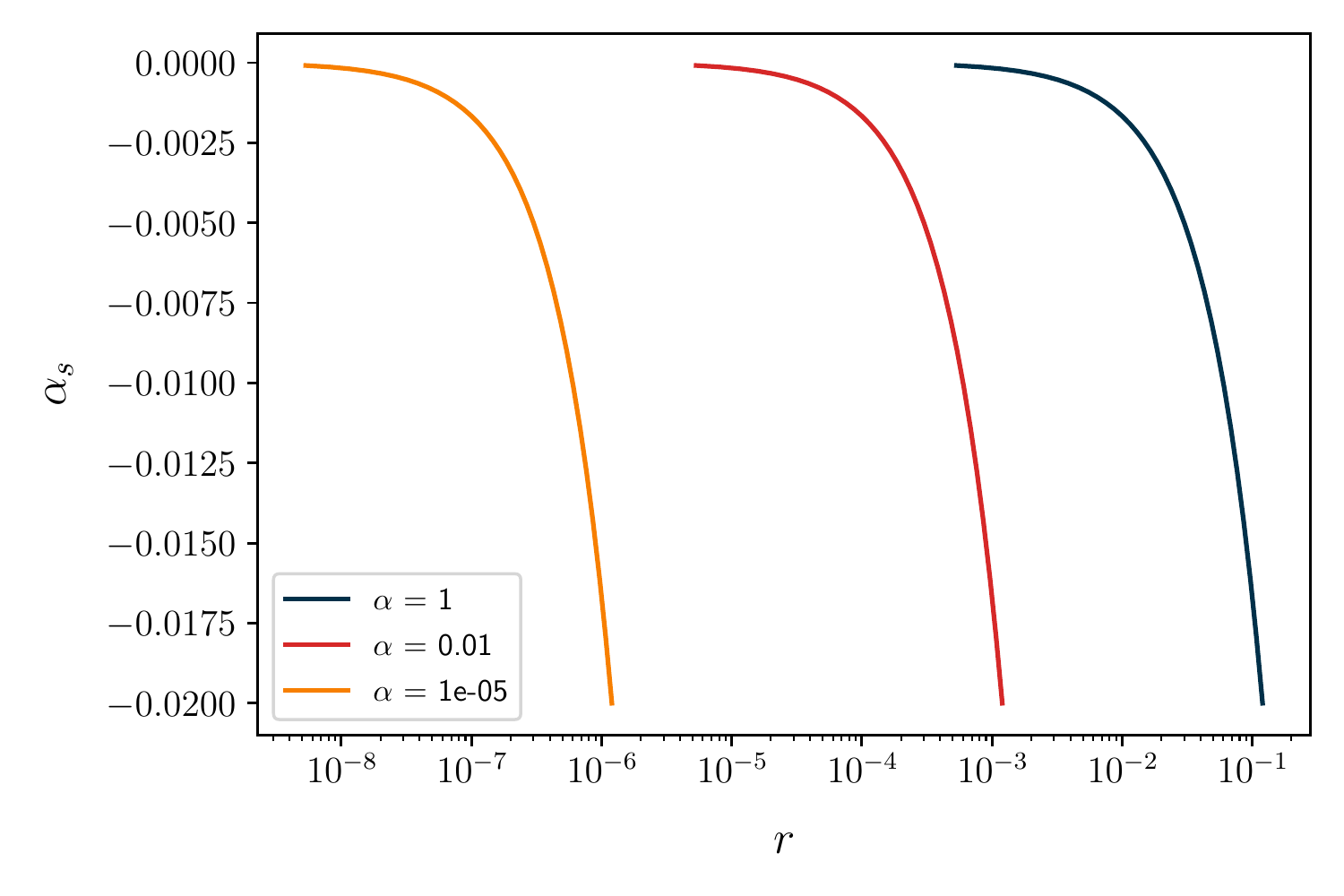}
    \caption{\textbf{Left}:  The scaling of the scalar index, $n_s$, as a funcion of the tensor-to-scalar ratio, $r$. \textbf{Right:} The scaling of the first order scalar running, $\alpha_S$, as a funcion of the tensor-to-scalar ratio, $r$. In both cases we show the scaling of the inflationary observables for three different values of $\alpha = 1,\ 0.01,\ 10^{-5}$.}
    \label{fig:potential}
\end{figure*}


In this paper, we discuss possible constraints of the model parameters using current CMB and LSS data in section \ref{sec.CMB_constraints} and in section \ref{sec.forecasts} we show possible constraints for the next generation of CMB experiments in the optimistic scenario of a not null measurement of primordial tensor modes. Finally, we summarize the main results, possible directions in future work and main conclusions.


\section{Constraints from current CMB and LSS data}\label{sec.CMB_constraints}


Current CMB \cite{Planck2013_inflat,Planck2015_inflat,BICEP,BICEP2,Aghanim:2018eyx} and LSS \cite{To:2020bhf} data are able to constrain the value of the scalar index $n_s$ with sub-percent accuracy, being the tensor amplitude $r \rightarrow 0$. Therefore, CMB Planck data are unable to constrain $\alpha$, and thus, sampling it in our analysis would give no insights on the $\alpha$-attractor models. Conversely, as we do here, one can fix the value of $\alpha$ and derive constraint on the inflationary parameters accordingly. Even though this approach is model-dependent, it may result in constraints that are not achievable when the standard approach is followed (in which any value compatible with the available data is permitted).
Moreover, when inflationary parameters are sampled independently from one another we can explore the full parameter space of the data. However, the parameter space of an inflationary model is just a subspace of this as it introduces correlations among the parameters that reduce the available parameter space. This is particularly evident if one considers the relation between $r$ and $n_s$ in $\alpha$-attractor model given by \autoref{eq:n_s}.

Along with the tensor-to-scalar ratio $r$, we consider as independent parameters the other five  standard $\Lambda$CDM ones: the baryon $\omega_b = \Omega_bh^2$ and the CDM $\omega_c = \Omega_c h^2$ densities, the Hubble constant $H_0$, the optical depth $\tau$ and the amplitude of scalar perturbations $A_s$. We also let free the running of the spectral index $\alpha_S$.  As $\alpha$-attractors satisfy the usual inflationary consistency relation, we fix the index of tensor modes to $n_T = -r/8$. The uniform prior distribution imposed on these parameters are reported in \autoref{tab:prior}. 

\begin{table}[h!]
\begin{tabular}{c|c}
\toprule 
\horsp
Parameter \vertsp Prior range\\
\hline
\hline
\morehorsp
$\Omega_bh^2$ \vertsp  $[0.005, 0.1]$  \\
\morehorsp
$\Omega_c h^2$ \vertsp  $[0.001, 0.99]$  \\
\morehorsp
$H_0$ \vertsp  $[20, 100]$ \\
\morehorsp
$\tau$ \vertsp $[0.01, 0.8]$ \\
\morehorsp
$r_{0.002}/\alpha\cdot10^{3}$ \vertsp $[0.5, 8] $ \\
\morehorsp
$\log(10^{10} A_s)$ \vertsp $[1.61, 3.91]$ \\
\morehorsp
$\sum m_\nu$ \vertsp $[0, 1]$ \\
\morehorsp
$N_\text{eff}$ \vertsp $[2, 5]$ \\
\bottomrule
\end{tabular}
\caption{Range of uniform priors distributions imposed on the sampled parameters during the analysis.}
\label{tab:prior}
\end{table}

Since CMB data bounds the scalar index $n_s \approx 0.96$ with sub-percent accuracy, the number of e-folds expected from $\alpha$-attractors (see \autoref{eq:N}) is $N\approx 60$. Therefore, selecting a value of $\alpha$ consequently fixes the order of magnitude of the tensor amplitude (see also \autoref{fig:potential}). To efficiently sample in the area of interest in the parameter space, the prior range for the tensor-to-scalar ratio $r$ is re-scaled appropriately with respect to the value of $\alpha$, so that their ratio is of the order of the value of $r$ expected for the Starobinsky model ($\alpha = 1$) \ie $r \approx 4\cdot 10^{-3}$.

The predictions of the theoretical observational probes are calculated using the latest version of the cosmological Boltzmann integrator code \texttt{CAMB} \cite{Lewis:1999bs,Howlett:2012mh}. To compare our theoretical predictions with data, we use the full 2018 Planck temperature and polarization datasets which also includes multipoles $\ell < 30$ \cite{Aghanim:2019ame}. We combine the Planck likelihood with the Biceps/Keck 2015 B-mode data \cite{Ade:2018gkx} and the combination of galaxy clustering and weak lensing data from the first year of the Dark Energy Survey (DES Y1) \cite{Abbott:2017wau}. The posterior distributions of the cosmological parameters have been explored using the publicly available version of the Bayesian analysis tool \texttt{COBAYA} \cite{torrado:2020xyz}. In particular, the posteriors have been sampled using the MCMC algorithm developed for CosmoMC \cite{Lewis:2002ah,Lewis:2013hha} and tailored for parameter spaces with a speed hierarchy.

\begin{figure*}
    \centering
    \includegraphics[width=.65\linewidth]{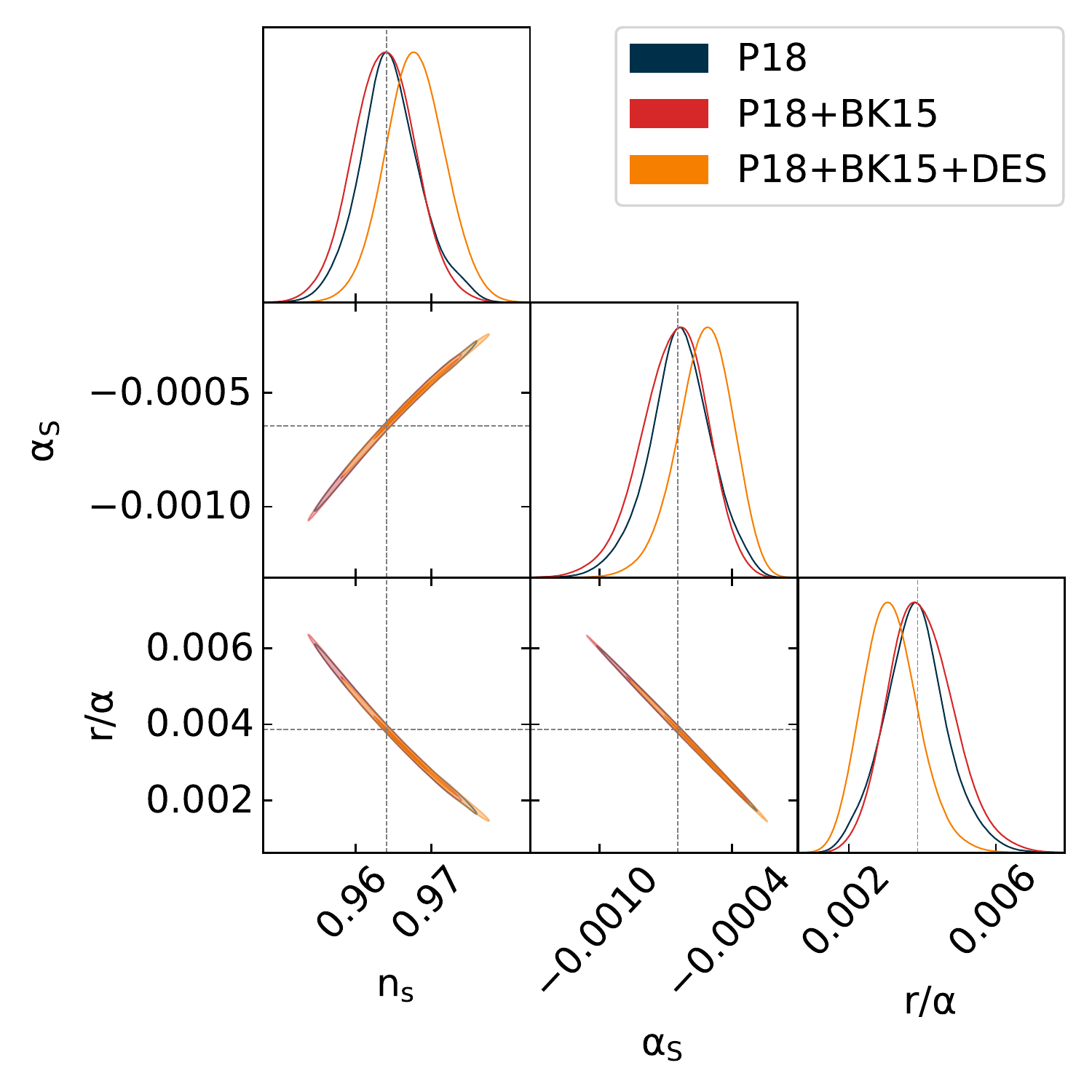}
    \caption{Posterior distributions and 2D contour levels at $68\%$ and $95\%$ C.L. for Planck 2018 data alone and combined with the Biceps/Keck 2015 B-mode data and with Large Scale Structure DES data. The dashed grey markers denote the expected values for Staronbinsky inflation ($\alpha = 1$ and $N\approx 60$).}
    \label{fig:planck_triangle}
\end{figure*}

The results of the Bayesian analysis are reported in \autoref{fig:planck_triangle}. Current CMB data are unable to constrain the value of $\alpha$ even imposing the $\alpha$-attractor model a priori. As previously discussed, CMB data are consistent with a vanishing tensor amplitude at the scale of the horizon exit of scalar perturbations ($k_\star \sim 0.05\ \rm Mpc^{-1}$) . In the context of $\alpha$-attractors, the tensor amplitude depends on the $\alpha$ parameter consequently the CMB constraint on the tensor amplitude is translated into a constraint on the ratio $r/\alpha$, \ie $r/\alpha \rightarrow 0$, which does not allow  for a measure of the value of $\alpha$.
However, the consistency relation of the $\alpha$-attractor model impose a correlation between the scalar index, $n_s$, with $r/\alpha$ (see~\autoref{eq:n_s}). Given that CMB data constraint the scalar index with sub-percentage accuracy ~\cite{Aghanim:2018oex}, this result in a measure of the ratio $r/\alpha$ in the context of the $\alpha$-attractors, $r/\alpha =  0.00387^{+0.00078}_{-0.00094}$. Interestingly this is the value of the tensor-to-scalar ratio expected in Starobinksy inflation for $n_s \approx 0.965$ or equivalently for $\alpha = 1 $ and $N \approx 60$.  
Due to this correlation, we see also that there is virtually no difference between the results using the Planck 2018 data and combining them with the Biceps/Keck 2015 data as the constraint on the scalar index is unchanged (if not for a statistically insignificant shift in the posterior mean between the two runs). Consequently, in Planck data, we found no correlation between $\alpha$ and the scalar spectrum parameters, $n_s$ and $\alpha_S$. When Large Scale Structure data (i.e: DES) is included in the analysis, a shift in the spectral index $n_s$ with respect to CMB data is found. DES data prefer a slightly higher value for $n_s$, shifting accordingly the running of the spectral index $\alpha_S$ and the tensor-to-scalar ratio $r$. However, the results remain consistent with that from PK18 and PK18+BK15 within $1\sigma$.

Incidentally, we also obtain a constraint on the running of the scalar index $\alpha_S$ (related to $r/\alpha$ by \autoref{eq:n_run}) away from zero at 4 standard deviation \ie $\alpha_S = -6.4^{+1.6}_{-1.3}\cdot 10^{-4}$. It is worth noting that this is not due to an indication of a scalar running in CMB data but to the specific correlation which arises in $\alpha$-attractor inflation between the parameters of the scalar and tensor spectrum. These results however point out either future measurements of $r_{0.002}$ and $n_{\rm nrun}$ could potentially rule out the $\alpha$ attractor model: they are key parameters in studying the viability of an inflationary model and should be considered in the future analysis of CMB and LSS data.

\section{Forecast for future CMB-S4 observations}\label{sec.forecasts}
We have also aimed to study the available parameter space for the $\alpha$-attractor model with future CMB experiments. In the previous section we have discussed the constraints achievable with current CMB and LSS data showing that, given the experimental sensitivity, current data are unable to constrain the value of the $\alpha$ parameter. However, future generation CMB experiments are expected to strongly improve the sensitivity on the B-mode polarization signal of the CMB, possibly discovering evidence of a primordial tensor mode with amplitude in the range of $r \sim 0.01 - 0.001$ \cite{Ade_2019, thesimonsobservatorycollaboration2019simons,Abazajian:2020dmr,Hazumi:2021yqq}. In particular, this is the range of predictions for many well-motivated inflationary models, such as Starobinsky inflation, considered the benchmark for future CMB observations \cite{Ade_2019, thesimonsobservatorycollaboration2019simons,Abazajian:2020dmr,Hazumi:2021yqq}.
In this section, we study the optimistic scenario of a future detection in the CMB anisotropies of non-vanishing tensor amplitude and we forecast the constraints achievable with a CMBS4-like experiment on the parameters of the $\alpha$-attractor model.

We consider as a baseline model a minimal extended $ \lcdm $ cosmology with the inclusion of non-vanishing tensor-to-scalar ratio, $r$ and $\alpha$. This extended model constitutes our simulated data sets. The value of $r$ is chosen correspondingly to the best-fit value obtained with a Starobinsky model using only Planck 2018 data \ie $r = 0.00387$, while we fix $\alpha = 1$. The value of the scalar index and its running are also fixed to $n_s = 0.964$ and $\alpha_S = 0.0006$. The remaining $\lcdm$ parameters values are: 
$\omega_b = 0.0221$, $\omega_c = 0.12$, $H_0 = 67.3$, $\tau=0.06$ and $\ln(10^{10} A_s) = 3.05$. As the new generation of CMB experiment also expects to set some light in the neutrino sector, we also explore the number of effective degrees of freedom of relativistic species $N_\text{eff}$ and the sum of the neutrino masses $\sum m_\nu$ in the forecast. 

Both simulated data and theoretical models are computed with the latest version of the Boltzmann code \texttt{CAMB} \cite{Lewis:1999bs,Howlett:2012mh}. To extract constraints on cosmological parameters, we make use of the Monte Carlo Markov Chain (MCMC) code CosmoMC \cite{Lewis:2002ah,Lewis:2013hha} which compares theory with a simulated dataset using a given likelihood.

\begin{figure*}[htp!]
   \centering
   \includegraphics[width=.5\textwidth]{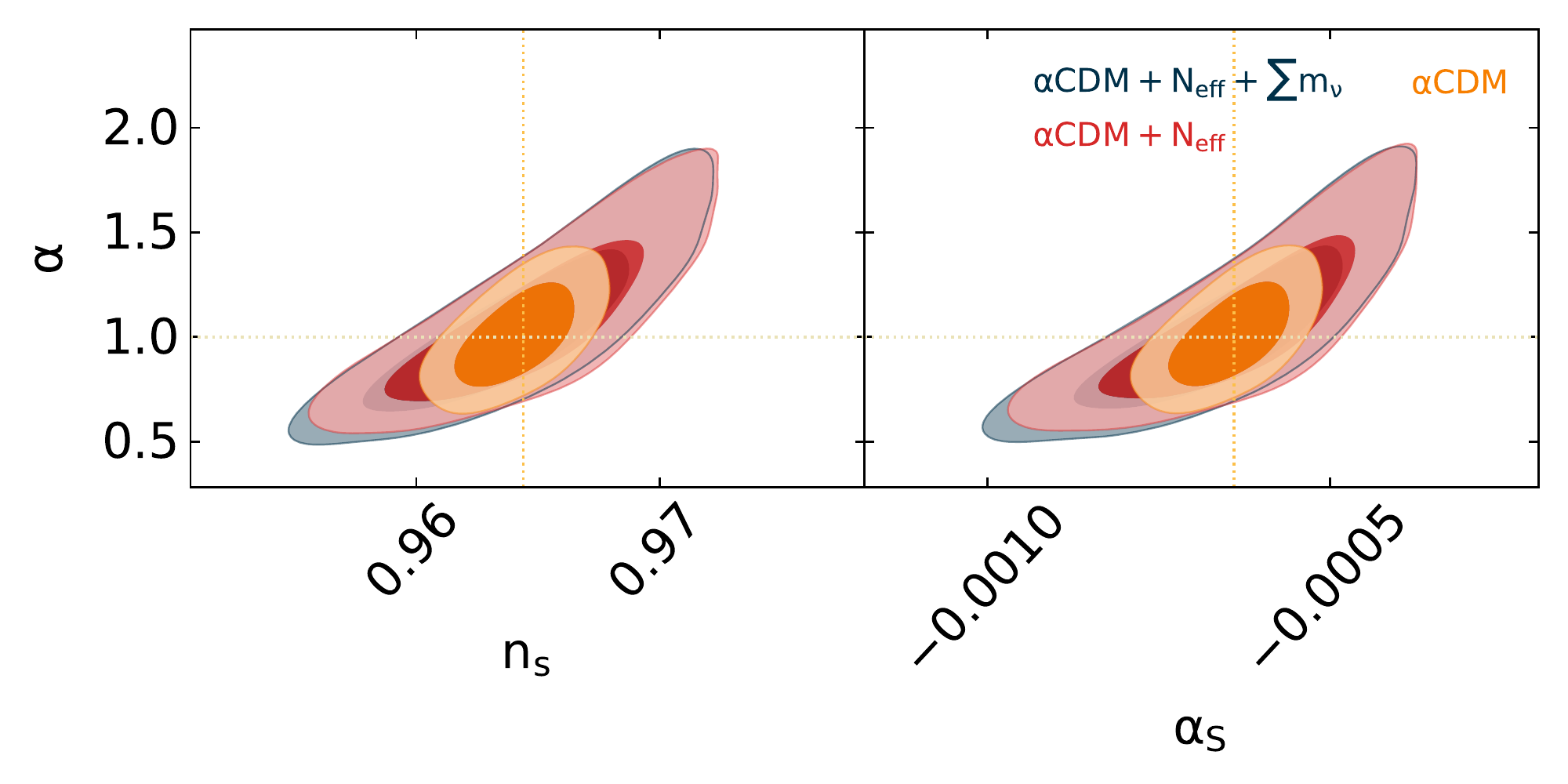}
   \includegraphics[width=.5\textwidth]{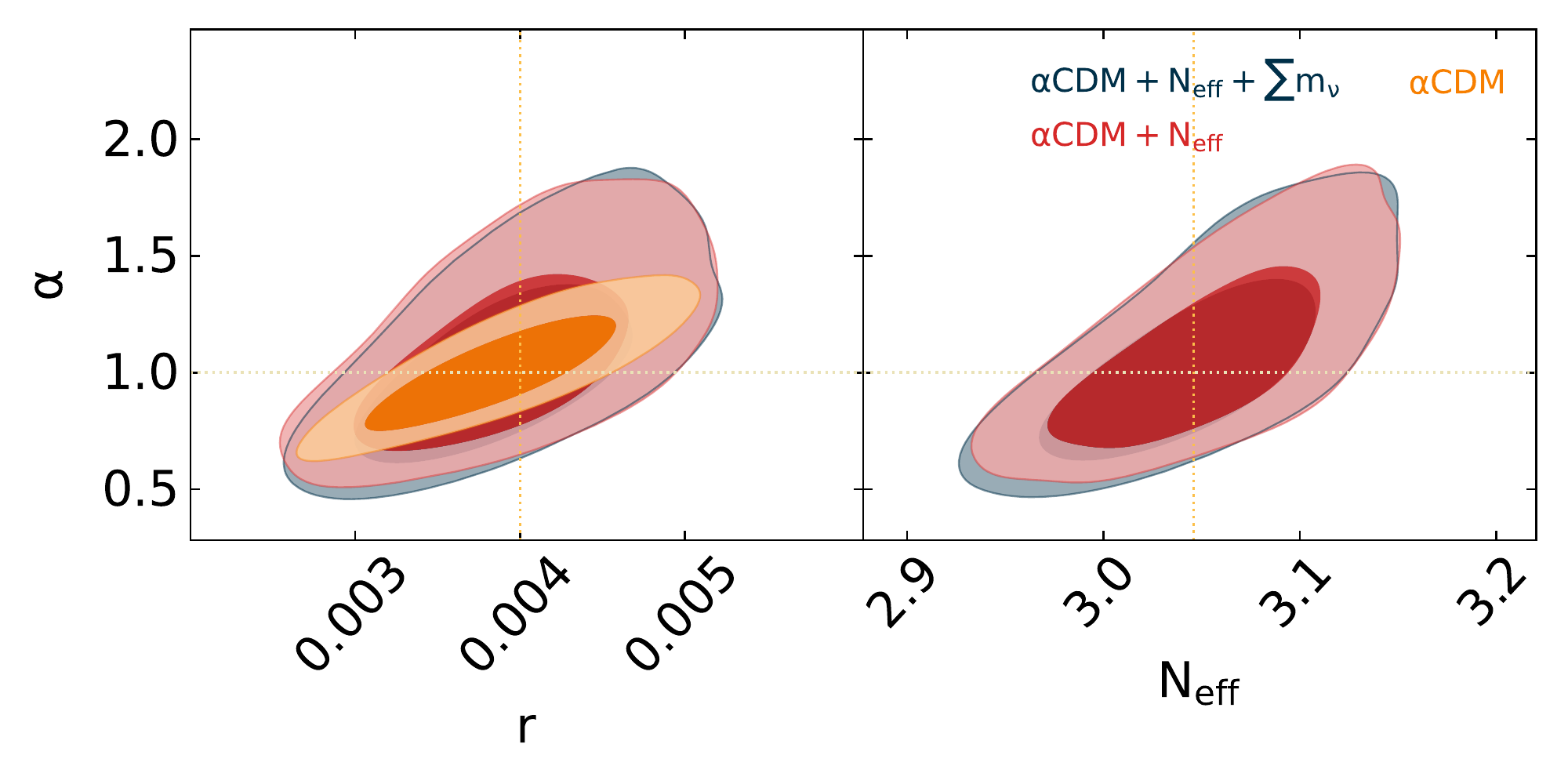}
   \caption{Forecasted 2D contours at $68\%$ and $95\%$ C.L. for $\alpha$ attractor inflationary parameters for a CMB-S4 experiments with and without allowing the neutrino sector to vary ($\alpha$CDM, $\alpha$CDM$+N_\text{eff}$ and $\alpha$CDM$+N_\text{eff}+\sum m_\nu$ respectively).}
   \label{fig:correlation_alpha}
\end{figure*}

As in \cite{DiValentino:2018jbh,Renzi:2017cbg,Renzi:2018dbq}, we built our forecasts for future CMB experiments following a well-established and common method. Using the set of fiducial parameters described above, we compute the angular power spectra of temperature $ C_\ell^{TT} $, E and B polarization $ C_\ell^{EE,BB} $ and cross temperature-polarization $C_\ell^{TE} $ anisotropies. Then, for a given experimental configuration, we consider a noise for the temperature angular spectra of the form \cite{Perotto:2006rj}:
\begin{equation}
	N_\ell = w^{-1}\exp(\ell(\ell+1)\theta^2/8 \ln 2)
\end{equation}  
where $ \theta $ is the FWHM angular resolution and $ w^{-1} $ is the experimental sensitivity expressed in $ \mu\mathrm{K}\,\rm arcmin $. The polarization noise is derived equivalently assuming $ w_p^{-1} = 2w^{-1} $ since one detector measures two polarization states. 
The simulated spectra, realized accordingly to the previous discussion, are compared with theoretical ones using the following likelihood $\mathcal{L} $ \cite{Perotto:2006rj,Cabass:2015jwe}:
\begin{equation}
	-2\ln\mathcal{L} = \sum_{\ell} (2\ell + 1)f_{sky}\left(\frac{D}{|\bar{C}|} + \ln\frac{|\bar{C}|}{|\hat{C}|} - 3 \right)
\end{equation} 
where $\hat{C} $ and $ \bar{C} $ are the theoretical and simulated spectra (plus noise), respectively and are defined by :
\begin{align}
&|\bar{C}| = \bar{C}_\ell^{TT}\bar{C}_\ell^{EE}\bar{C}_\ell^{BB} -
\left(\bar{C}_\ell^{TE}\right)^2\bar{C}_\ell^{BB}  \\
&|\hat{C}| = \hat{C}_\ell^{TT}\hat{C}_\ell^{EE}\hat{C}_\ell^{BB} -
\left(\hat{C}_\ell^{TE}\right)^2\hat{C}_\ell^{BB}
\end{align}
while $ D $ is defined as:
\begin{align}
	D  &=
	\hat{C}_\ell^{TT}\bar{C}_\ell^{EE}\bar{C}_\ell^{BB} +
	\bar{C}_\ell^{TT}\hat{C}_\ell^{EE}\bar{C}_\ell^{BB} +
	\bar{C}_\ell^{TT}\bar{C}_\ell^{EE}\hat{C}_\ell^{BB} \nonumber\\
	&- \bar{C}_\ell^{TE}\left(\bar{C}_\ell^{TE}\hat{C}_\ell^{BB} +
	2\hat{C}_\ell^{TE}\bar{C}_\ell^{BB} \right). \nonumber\\
\end{align}

For this paper, we have constructed synthetic realizations of CMB data for only one experimental configuration, namely CMB-S4 (see e.g. \cite{Abazajian:2016yjj}). The CMB-S4 dataset is constructed using $ \theta = \SI{3}{\arcminute} $ and $ w = 1\, \si{\mu\kelvin}\,\rm arcmin $, and it operates over the range of multipoles  $ 5 \leq \ell \leq 3000 $, with a sky coverage of the $ 40\% $. Furthermore CMB-S4 is expected to reach a target sensitivity on the tensor-to-scalar ratio of $\Delta r \sim 0.0006$, whose goal is to provide a $95\%$ upper limit of $r < 0.001$. Therefore the value chosen for our fiducial model is well within the scope of an experiment like CMB-S4. However, the corresponding sensitivity on the value of the running of the scalar index, $\alpha_S$ would be only $ \Delta\alpha_S =  0.002$ which would clearly not be enough for a joint detection of $r$ and $\alpha_S$ assuming Starobinsky inflation. Thus, it may not be possible to distinguish between a generic $\alpha$-attractor model with $r \sim 0.004 $ and Starobinsky inflation despite a future detection of a non-vanishing tensor amplitude.  

In $\alpha$-attractors, however, the uncertainties about the correct shape of the inflationary potential, defining the value of $r$, $n_s$ and $\alpha_S$, are parameterized with the $\alpha$ parameter (see also \autoref{fig:potential}).  Therefore a measure of the value of $\alpha$ would also give us insights about the correct shape of the inflationary potential and correspondingly on the correct theory of inflation. 
A CMBS4-like experiment will be able to give such insights provided a detection of a non-vanishing tensor amplitude. Current data only place a loose upper bound $0 \leq \alpha \lesssim 10$ \cite{Kallosh:2019hzo} correspondingly to P18+BK15 upper limit on the tensor amplitude $r < 0.056 $ at $95\% $C.L. \cite{Akrami:2018odb}, see \autoref{eq:n_s}. To correctly explore the available parameter space for $\alpha$ we therefore employ a logarithmic prior on its value $ -6  \leq \log_{10}\alpha \leq 1 $ while we keep the priors on the other parameters as of \autoref{tab:prior}. We refer to this model as $\alpha\text{CDM}$. 
From our CMB-S4 forecasts we obtain a $15\%$ bound on the parameter $\alpha = 1.01 ^{+0.14}_{-0.18}$, clearly showing the ability of future CMB experiments of bounding single-field slow-roll inflationary models. Models with $\alpha \geq 2 $ and $\alpha \leq 0.5$ would be potentially excluded at more than 2 standard deviations in the optimistic scenario of a PGWs detection with amplitude in the range of the Starobinsky model. We eventually extend this baseline model including the number of relativistic neutrino species $N_\text{eff}$, ($\alpha\text{CDM}+N_\text{eff}$). When  $N_\text{eff}$ is varied, we find a $5\%$ reduction of the accuracy with which $\alpha$ is measured \ie $\alpha = 1.07 ^{+0.18}_{-0.23}$ while the bound on the tensor-to-scalar ratio is basically the same in the two cases, \ie $\sigma(r) = 0.00050 $ . Conversely we found an increase in the error budget of the scalar index and running, passing from $\sigma(n_s) = 0.0016$ and $\sigma(\alpha_S) = 0.00006$ ($\alpha\text{CDM}$) to  $\sigma(n_s) = 0.0035$ and $\sigma(\alpha_S) = 0.0001$ ($\alpha\text{CDM}+N_\text{eff}$) an worsening of a factor around two in both cases. It is worth stressing that Primordial Gravitational waves may also contribute to the number of relativistic species being themselves relativistic degrees of freedom \cite{Cabass:2015jwe,Smith:2006nka,Clarke:2020bil}. This contribution can be calculated analytically to be :
\begin{equation}
    N_\text{eff,GW} \sim \frac{rA_s}{n_T}(A^{n_T} - B^{n_T})
\end{equation}
where $A$ and $B$ are two real numbers and $A,B \gg 1$. This contribution is clearly extremely small for red spectra ($n_T \leq 0 $) but may be important in inflationary theories where blue spectra ($n_T > 0$) can be produced (see e.g. \cite{Mukohyama:2014gba,Namba:2015gja,Stewart:2007fu,Ozsoy:2020ccy,Peloso:2016gqs,Giare:2020vhn,Giare:2020vss}). Consequently the only interaction between PGWs and neutrinos considered in this work is the one arising from neutrino anisotropic stress after  neutrino decoupling at $T \lesssim 1\ \text{MeV}$ \cite{Kojima:2009gw}. 
These constraints are virtually unmodified when we further extend our baseline model, allowing the whole neutrino sector to vary \ie $N_\text{eff} + \sum m_\nu$.
The 2D contours for both our forecasts are reported in Fig.\ref{fig:correlation_alpha}. A strong correlation now arises between $\alpha$ and the other inflationary parameters conversely to what we found with the Planck data. This is due to the power of CMB-S4 of resolving the B-mode spectrum, consequently breaking the degeneracy between $r$ and $n_s$. 
Nevertheless, the situation is unchanged for the scalar running. The strong bound we find on the scalar running is in fact due to imposing the $\alpha$-model a priori. Even a StageIV experiment would not have the required accuracy to measure the tiny scalar running predicted by $\alpha$-attractor inflation.  When $r$, $n_s$ and $\alpha_S$ are independently varied (\ie neglecting the consistency relation in \autoref{eq.infparameters_obs}) the running is fixed only with an error $\sigma(\alpha_S) = 0.0029$ at $68\%$ C.L., an order of magnitude higher than when the $\alpha$-model is imposed \emph{a priori} and in good agreement with the expected sensitivity for the CMB-S4 experiment \cite{CMB-S4}.

\section{Conclusions}
We have carried out a Bayesian analysis with current CMB and LSS data to constrain inflationary observables (the scalar spectral index $n_s$, its running $\alpha_S$ and the tensor-to-scalar ratio $r$). It has been observed that, with the current constraining power on $n_s$ and imposing the $\alpha$-attractor model a priori in our analysis, the possible values of the ratio $r/\alpha$ are narrowed in a band of around 0.004.  We conclude that current data do not have enough sensitivity to constrain any deviation from a Starobinsky inflationary model as the predicted tensor-to-scalar ratio is much smaller than the current upper limit of LSS and CMB data \ie $r \ll 0.1$. 

Later, we turned our attention to future CMB probes, focusing on a CMB StageIV-like experiment and using the constraints from current Planck data as a benchmark for our forecast. This led to a mock dataset with $r \neq 0$ and we fixed $\alpha = 1$ effectively reproducing Starobinsky inflation.

The forecast is performed from a Bayesian statistical approach, where $\alpha$ is let free to be sampled from a logarithmic prior distribution. Future CMB-S4 experiments will then be able to constrain $\alpha$ as long as the value of $r$ is above the target sensitivity expected from such experiment \ie $r > 0.001$. Conversely, in the pessimist scenario that even in future CMB-S4 data a vanishing tensor-to-scalar ratio will be measured, the situation will be exactly as for current data and only an upper limit on the value of $\alpha$ could be placed. We can forecast the corresponding upper limit on $\alpha$ from \autoref{eq:n_s}. Assuming $n_s = 0.9644$ and $r \leq 0.001$ one finds $\alpha \lesssim 0.26$ which would lead to excluding many inflationary models, in particular, the Starobinsky model would be excluded at more than six standard deviations. 

In conclusion, a future CMB-S4 experiment will have enough sensitivity to significantly constrain single-field slow-roll inflationary models. In the case of an optimistic detection of a non-vanishing tensor amplitude, it would be able to shed light on both the energy scale and the shape of the inflationary potential, while in the pessimistic scenario of a non-detection of tensor modes it would still be able to place a tight upper limit on the value of $\alpha$ and exclude Starobinsky inflation at more than $6\sigma$ .
We underline that, when the running of the spectral index $\alpha_S$ is free to vary, it is always different from zero as expected from the inflationary consistency relation of the $\alpha$ attractor model. However, we show that the value expected for the scalar running given the current constraints on the scalar index is so small that it will not be detectable by a future CMB-S4 experiment (with an expected sensitivity of $\Delta \alpha_S \sim 0.003$), but it may be reachable when information from future weak lensing and galaxy clustering measurements will be included \cite{Blanchard:2019oqi,Font-Ribera:2013rwa,LSST_sciencebook}.

\acknowledgements
We thank Ana Ach\'ucarro, William Giarè and Santiago Casas for useful comments and discussion. GCH acknowledges support from the Delta Institute for Theoretical Physics (D-ITP consortium), a program of the Netherlands Organization for Scientific Research (NWO). FR also acknowledges support from the NWO and the Dutch Ministry of Education, Culture and Science (OCW), and from the D-ITP consortium, a program of the NWO that is funded by the OCW.

\bibliographystyle{aipnum4-1}
	
\bibliography{biblio}
\end{document}